\documentstyle[12pt,worldsci]{article}
\def\la{\mathrel{\mathpalette\fun <}}

\def\fun#1#2{\lower3.6pt\vbox{\baselineskip0pt\lineskip.9pt
  \ialign{$\mathsurround=0pt#1\hfil##\hfil$\crcr#2\crcr\sim\crcr}}}

\begin{document}

\title{{\bf LIMITS ON NEUTRINO RADIATIVE DECAY FROM SN1987A}}
\author{ANDREW H. JAFFE\\
{\em Enrico Fermi Institute and Department of Astronomy \& Astrophysics,\\
University of Chicago, 5640 S. Ellis Ave., Chicago, IL 60637-1433, USA}\\
\vspace*{0.3cm}
ED FENIMORE\\
{\em MS D436, Los Alamos National Laboratory, Los Alamos, NM 87544, USA}\\
\vspace{0.3cm}
and\\
\vspace*{0.3cm}
MICHAEL S. TURNER\\
{\em Enrico Fermi Institute and Departments of Physics and
 Astronomy \& Astrophysics,\\
University of Chicago, 5640 S. Ellis Ave., Chicago, IL 60637-1433, USA\\
and\\
NASA/Fermilab Astrophysics Center,\\
Fermi National Accelerator Laboratory, Batavia, IL 60510-0500, USA}
}

\maketitle
\setlength{\baselineskip}{2.6ex}

\begin{center}
\parbox{13.0cm}
{\begin{center} ABSTRACT \end{center}
{\small \hspace*{0.3cm}
 We calculate limits on the properties of
neutrinos using data from gamma-ray detectors on the Pioneer Venus
Orbiter and Solar Max Mission satellites. A massive neutrino decaying in
flight from the supernova would produce gamma rays detectable by these
instruments. The lack of such a signal allows us to constrain the mass,
radiative lifetime, and branching ratio to photons of a massive neutrino
species produced in the supernova.}}
\end{center}

\section{Introduction}
The occurence of Supernova 1987a in the Large Megellanic Cloud has
proven to be among the most fruitful experiments in the heavenly
laboratory for the confirmation of ``known'' physics and the constraining of
new physics. Aside from its obvious impact upon the study of the late
stages of stellar evolution in general and upon supernova physics in
particular, models for SN87a have become an industry for the study of
the couplings of light particles (neutrinos, axions) to ordinary matter.
In this work, we discuss limits upon the properties of neutrinos
independent of specific model for the supernova.

When a supernova occurs, the bulk of the binding energy of the
progenitor star ($\sim3\times10^{53}$ erg) is released in neutrinos, a
fact predicted by theory and confirmed by the observation of a neutrino
burst from the supernova, with a characteristic temperature of about
$T_\nu\approx4.5$ MeV. If at least one species of neutrinos is
unstable and if it couples to the
photon, then some of these neutrinos will decay to photons en route,
which are potentially
detectable as MeV gamma rays. At the time of the supernova burst's
arrival at earth and environs, there were several satellites operating
in the solar system capable of detecting the decay photons in the course
of their watch for gamma ray bursts. Analyses of the data from one of
these detectors, on board the Solar Max Mission (SMM) satellite, has
already been presented\cite{1,2,3,4}. Here, we examine the data from Gamma
Burst Detector on the Pioneer Venus Orbiter (PVO).

\section{Expected Gamma-Ray Signal}

We assume that $1/3$ of the total supernova energy is released in a
species of massive neutrinos. These neutrinos then decay in flight into
a photon and a light neutrino ({\it e.g.}, an electron neutrino). Typically,
the decay products will each have energies of
$1/2 \times 3 \times T_\nu \approx 7$ MeV.

We will consider decays of the form $\nu\rightarrow\nu'+\gamma$. We
expect that the parent neutrino will be a massive exotic neutrino, while
the daughter neutrino will be a member of a light family such as
$\nu_e$. In particular, this decay allows two possibilities: the
helicity of the daughter neutrino may either be flipped or the same with
respect to the parent as the photon takes away a unit of spin (assuming
both neutrinos are relativistic). From quantum mechanics, then, we know
that the distribution of the photon in the rest frame of the
parent will be proportional to either $\cos^2(\bar\phi/2)$ or
$\sin^2(\bar\phi/2)$, where $\bar\phi$ is the rest-frame angle between
the directions of the parent neutrino and the photon.

At a given time at the detector, photons are received that resulted from
neutrinos decaying on the surface of an ellipsoid defined such that the
sum of the time between the supernova and the neutrino decay and the
time between the decay and the detection of the photon is equal to the
time of detection. Actually, this is only approximately true---there is
an additional delay incurred due to the finite mass of the parent
neutrino. Furthermore, due to the relativistic beaming of the decay
products into the forward direction, many more photons are received from
long, skinny ellipsoids (corresonding to short delay times) than from
others (corresponding to longer delays). Thus, the relativistic delay
will play a correspondingly larger role in these events.

{}From relativistic kinematics, the angular distribution of daughter
photons in turn determines the distribution of photon energies as a
function of neutrino energy, for each of the two possible decays (flip
or no flip). The time delay is determined by the energy of the neutrino
and the decay ellipsoid above.
Thus, the photon spectrum as a function of time is determined. It is
given by
\begin{eqnarray}
dN &=& B_\gamma
{L_\#(E)\over4\pi D^2}{e^{-t_d/\gamma\tau}\over\gamma\tau}
f_i(E, \mu) \,dt\,dt_d\,dE\,d\mu \nonumber \\
& & \mbox{}\times \delta\left(t-t_d\left\{1-v\mu+{D\over t_d}
   \left[\sqrt{1-\left(vt_d\over D\right)^2(1-\mu^2)}-1\right]\right\}\right)
\end{eqnarray}
where $B_\gamma$ is the branching ratio of the parent neutrino to photons,
$N$ is the neutrino number flux at the detector, $L_\#$ is the
differential number luminosity as a function of E, the parent neutrino
energy, $D=1.7\times10^{23}$cm is the distance to the supernova, $t_d$
is the time of the neutrino's decay, $v$ is the neutrino's velocity and
$\gamma=E/m_\nu$ is the relativistic factor, $\tau$ is the neutrino lifetime,
$t$ is the time of the photon's arrival at the detector, $\mu$ is the
cosine of the ``lab-frame'' decay angle between the parent neutrino's
direction and the photon, and $f_i(E,\mu)$ is the distribution of angles
as a function of neutrino energy for each of the two helicity
possibilities. The Dirac $\delta$ function enforces the arrival
ellipsoid. In this paper, we will concentrate on the limit of short
decay times, $t_d\sim\gamma\tau \ll D$, so the delta function becomes
$\delta\left(t-t_d(1-v\mu)\right)$.

Changing variables from $\mu$ to the photon energy $k$, and integrating over
the decay time $t_d$ gives
\begin{equation}
dN = B_\gamma
{L_\#\over4\pi D^2} e^{-2kt/m_\nu\tau} {2k\over m_\nu\tau}f_i(E,k)
\,dk\,dE\,dt,
\end{equation}
where $f_i(E,k)$ now gives the distribution of photon energies as a
function of parent neutrino energy:
\begin{equation}
f(E,k) = \cases {
2k/E^2 &no flip\cr
2(E-k)/E^2 &flip.\cr
}\end{equation}
Similar results have been derived for related cases before\cite{5,6,7}.
Note that this distribution is a function only of the combination of
neutrino parameters $m_\nu\tau$ and $B_\gamma$; to break this degeneracy
between $m_\nu$ and $\tau$, we must relax our assumption of short decay
times.

We will also assume that the initial neutrino luminosity is given by
a zero-chemical-potential Fermi-Dirac distribution. Normalized to the
known temperature and total energy of the suprenova, this gives
\begin{equation}
L_\#(E) =
{120\over7\pi^4}{E_T\over T_\nu^4} {E^2\over1+e^{E/T_\nu}}
.\end{equation}

Finally integrating over this distribution, this gives the photon
spectrum as a function of time,
\begin{equation}
{dN\over dk\,dt} =
{240\over 7\pi^4} {B_\gamma \over 4\pi D^2} {E_T\over T_\nu^2}
{2k\over m_\nu\tau} e^{-2kt/m_\nu\tau} h_i(k/T_\nu).
\end{equation}
where $h_i$ is a separate function for each of the helicity
possibilities that is of order unity for the energies and temperatures
of interest.

\section{PVO Instrumentation}

The Pioneer Venus Orbiter Gamma Burst Detector (OGBD)\cite{8} was
designed to detect gamma ray bursts---transient, high energy events that
last from milliseconds to tens of seconds. It has four separate bands,
0.1--0.2 MeV, 0.2--0.5 MeV, 0.5--1 MeV, and 1--2 MeV. In the background
mode that we will be using, it has a timing resolution of 12 or 16
seconds with full spectral information ({\em i.e.}, counts in each of
the four bands).

Because gamma-ray bursts are singular events, the OGBD was designed to
have full-sky coverage, although it is incapable of independently
providing directional information about the photons it receives.
However, the OBGD detectors do, in fact, face the South Ecliptic Pole,
and thus are ideally suited for measurements of gamma-rays from the LMC.
Thus, even though it was not ``pointing'' at the LMC at the time, it can
still be used to analyze data associated with the supernova. Although
the response of the instrument changes for is a function of the
gamma-ray direction, the bulk of the decay photons come from the
direction of the supernova itself on the sky (the exceptions are the
rare photons which reach us after decaying at a very large angle from
the outgoing neutrino).

\section{Analysis}

After a brief look at the data from the four OGBD channels in Fig. 1,
it is clear that there is no obvious signal of gamma rays over the
background (the variation in the signal is in fact consisitent with
pure $\sqrt{N}$ noise). Thus, we will be putting limits on our free
parameters: $m_\nu\tau$ and $B_\gamma$.

\vspace*{11cm}
\begin{center}
{\small Fig. 1.  Raw counts/sec in each of the four OGBD
channels for the time surrounding the arrival at Venus of the
supernova's optical burst, at time $UT-27325=0$ sec.}
\end{center}

To do this, we utilize the SPANAL code which was developed for analyzing
the instantaneous spectra of gamma-ray bursts.  Given the spectrum
averaged over 12- or 16-second bins, SPANAL calculates the best fit set
of parameters, $m_\nu\tau$ and $B_\gamma$, for the data (thus far, we
simulataneously fit the data for up to six time bins, or about 90
seconds after the supernova). In Fig. 2, we show the results for a small
area of parameter space. The points are $\chi^2$ minima, with
$\chi^2\approx22$ (with 2 degrees of freedom) for all of the points. For
the point $m_\nu\tau=10^3$ keV sec, we show in addition the 1-$\sigma$
upper limit on $B_\gamma$; it is not very far from the $\chi^2$ minimum,
so we expect that the actual upper limits are not far from these points
throughout.  The area below the line is roughly excluded; the area above
is allowed by the OBGD data. To compare with previous analyses, Kolb and
Turner\cite{2} found a limit in a similar area of parameter space of
$B_\gamma\la2.8\times10^{-10}$ without taking into account the angular
distribution of decay photons (by simply assuming $k=E/2$).

In order to more fully map the limits of parameter space, we must keep
the full information of the arrival ellipsoid in the delta function of
Eq. 1. For such long arrival times that this effect matters,
however, the expected number of gamma rays is much lower (since fewer
neutrinos would have decayed before reaching earth), so the limits will
be somewhat weaker. We are currently pursuing the analysis in this
realm.

\vspace*{11cm}
\begin{center}
{\small Fig. 2.  $\chi^2$ minima for the fit of the theoretical
spectrum of photons (Eq. 5) to the OGBD data after the
supernova. For $m_\nu\tau=10^3$ keV sec, we also show the
1-$\sigma$ upper limit.}
\end{center}

\bibliographystyle{unsrt}

\end{document}